\documentstyle[prl,aps,epsf, amsfonts]{revtex}

\def\8{\infty}

\def\undertext#1{\vtop{\hbox{#1}\kern 1pt \hrule}}

\def\VEV#1{\left\langle\,#1\,\right\rangle}

\def\br{\\ \nonumber & &}

\def\be{\begin{equation}}
\def\ee{\end{equation}}
\def\bea{\begin{eqnarray} & &}
\def\eea{\end{eqnarray}}
\def\ct#1{\cite{#1}}
\def\rf#1{(\ref{#1})}

\title{c-Theorem for Disordered Systems}

\author {V. Gurarie }

\address{Institute for Theoretical Physics, University of California,
Santa Barbara CA 93106-4030}

\date{\today}

\begin {document}
\draft
\maketitle

\begin{abstract}

We find an analog of Zamolodchikov's c-theorem for disordered 
two dimensional noninteracting systems in their supersymmetric
representation. For this purpose we introduce a new parameter
$b$ which flows along the renormalization group trajectories much
like the central charge for unitary two dimensional field theories. However,
it is not known yet if 
this flow is irreversible.
$b$ turns out to be related to the central extension of a certain algebra,
a generalization of the Virasoro algebra, which we show may be
present at the critical points of these theories.  $b$ is also 
related to the physical free energy of the disordered system defined
on a cylinder. We discuss possible applications by computing $b$
for two dimensional
Dirac fermions with random gauge potential.

\end{abstract}

\vspace{.5in}
A basic fact in the theory of disordered noninteracting systems is that 
they can
be described by the special supersymmetric field theories, the
theories which  
contain equal number of bosonic and fermionic fields
and which are
invariant under {\sl supergroups}, the groups which rotate
bosonic and fermionic fields into each other \ct{Efetov,Bernard}. 
Unfortunately it is generally very hard to solve these theories. 
Perturbative methods remain the only tool used consistently on them. 
One of the main difficulties
in treating these theories stems from the fact that they are nonunitary,
having been constructed in such a way as to make their partition function 
equal to 1. 

Nevertheless there is certain hope that in two dimensions these theories
must be treatable. After all, they are not much different from other
two dimensional field theories whose critical points are described by
conformal field theory \ct{BPZ}. It goes without saying that solving these
theories is very important for comparison with experiment \ct{Efetov}. 

It can be shown (see below) 
that if a conformal field theory describes a critical point
of one of these supersymmetric theories, its central charge $c$  
must be equal
to 0. There is a large number of conformal field theories whose central
charge is equal to 0, the direct product of two conformal field theories with
central charge equal to an arbitrary number 
$c$ and to $-c$ being just one example. Therefore,
the central charge loses its meaning as a way to distinguish between
different critical points of these theories. 

In this paper we introduce a parameter $b$ which replaces the central
charge as a way to distinguish between different critical points of
these supersymmetric theories. $c$ is equal to zero
because it counts the difference between bosonic and fermionic
degrees of freedom, which is 0 for those theories. Roughly speaking $b$
counts the {\sl sum} of bosonic and fermionic degrees of freedom. 
Moreover, we show that $b$ flows along the renormalization group 
trajectories in a way similar to $c$ in unitary theories, 
as expressed in the so-called
c-theorem \ct{cZ}.

A generic supersymmetric disordered field theory
is invariant under the group which
mixes bosons and fermions. We will choose this group to be
U$(1|1)$, even though the actual symmetry group of our theory could be
larger or could be based on other groups \ct{Efetov,Bernard}. 
We will denote the four generators of this group by $J$, $j$, 
$\eta$ and $\bar \eta$, following \ct{CMW}. 
As discussed in appendix \ref{apone} the action of such a field theory is
a part of the supermultiplet consisting of $S$, $s$, $\zeta$,
and $\bar \zeta$, the first two being commuting and the second two
anticommuting quantities. As a part of the supermultiplet 
the action itself is 
a variation of either $\zeta$ or
$\bar \zeta$ under the action of either $\bar \eta$ or $\eta$ respectively.
Correspondingly the energy-momentum tensor $T_{\mu \nu}$
is also a part of the supermultiplet consisting of
$T_{\mu \nu}$, $t_{\mu \nu}$, $\xi_{\mu \nu}$ and $\bar \xi_{\mu \nu}$. 
Because of this, the energy-momentum tensor is a variation of either
$\xi$ or $\bar \xi$ under the action of either $\bar \eta$
or $\eta$ respectively. 

The theory we are considering can
be thought of as a topological field theory. The topological
field theories are defined as theories with expectation values of
the products of any number of 
the energy momentum tensors equal to 0, and indeed for our theory
\be
\label{expec}
\VEV{T_{\mu \nu}(x) T_{\mu' \nu'}(x') \dots} = \delta_{\bar \eta}
\VEV{\xi_{\mu \nu}(x) T_{\mu' \nu'}(x') \dots } = 0
\ee
which is consistent with the fact that the partition function 
of this theory is equal to 1. 

Now it follows from \rf{expec} that if such a field theory
has a critical point, its central charge will be equal to zero
because the central charge is related to the expectation value
of the products of energy momentum tensors \ct{BPZ}. In fact, 
Zamolodchikov's $c$-function \ct{cZ}, which is equal to the central
charge at the critical point and usually changes along the renormalization
group trajectories, will be equal to $0$ everywhere. 
Therefore, the central charge is not a useful way to parametrize 
such a field theory. 

However, the correlation function of the energy momentum tensor 
$T_{\mu \nu}$ with $t_{\mu \nu}$ does not have to be zero. At this point,
following Zamolodchikov \ct{cZ}
we introduce the following notations
\be
\label{defcor}
\VEV{T(z, \bar z) t(0,0)} = {F(z \bar z) \over z^4}, \ \
\VEV{T(z, \bar z) \theta(0,0)} = {G(z \bar z) \over z^3 \bar z}, \ \
\VEV{\Theta(z, \bar z) \theta(0,0) } = {H(z \bar z) \over z^2 \bar z^2}
\ee
where $T\equiv T_{z z}$, $\Theta\equiv T_{z \bar z}$ and
$t\equiv t_{z z}$, $\theta\equiv t_{z \bar z}$. 

The correlation functions \rf{defcor} (and their complex conjugate 
counterparts) exhaust all the possible correlations between $T_{\mu \nu}$
and $t_{\mu \nu}$ by virtue of the following remarkable identity
\be
\label{identity}
\VEV{T_{\mu \nu}(x) t_{\mu' \nu'}(0)} = \VEV{t_{\mu \nu}(x) T_{\mu' \nu'}(0)}
.\ee
Its proof is discussed in the 
appendix \ref{aptwo}. It allows to relate various
correlation functions, for example, 
$\VEV{T(z, \bar z) \theta(0)} = \VEV{\Theta(z, \bar z) t(0)}$.

Now the tensors $\xi_{\mu \nu}$, $\bar \xi_{\mu \nu}$ and $t_{\mu \nu}$
are not necessarily conserved. However, combining the fact that
$T_{\mu \nu}$ is conserved with the identity \rf{identity}
we can show that
\be
\label{conserv}
\VEV{\partial_\mu t_{\mu \nu}(x) T_{\mu' \nu'}(0) } =
\VEV{\partial_\mu T_{\mu \nu}(x) t_{\mu' \nu'}(0) } =
0
.\ee
Therefore, as far as the correlators \rf{defcor} are concerned, we can work
with $t$ as if it were a conserved tensor. 
This allows us to derive the analog of Zamolodchikov's $c$-theorem.

We define
\be
b=F-2G-3H
.\ee
Now
we are in the position to claim that the following relation is true,
\be
\label{ctheorem}
{d b(t) \over d t} = - 6 H
\ee
where $t$ is a parameter which increases along the renormalization group
trajectory along the infrared direction. Indeed, 
Zamolodchikov considered a function $c=2F'-4G'-6H'$ where $F'$, $G'$,
and $H'$ are the correlators different from \rf{defcor} only by replacing
$t$ by $T$ everywhere. He showed that $dc/dt=-12H'$. In his proof
the only relevant information was that 
$T_{\mu \nu}$ was conserved. Fortunately
by virtue of \rf{conserv} we can work with $t_{\mu \nu}$ as if it were
conserved and arrive at \rf{ctheorem} following Zamolodchikov's
work step by step.

Does the parameter
$b$ decrease along the renormalization group trajectories? 
We cannot give a definite answer to this question at this time, but we find
the following heuristic arguments useful. 

$b$ does decrease  along the renormalization group trajectories
as long as
$H>0$. 
Can we show that $H>0$? As shown in appendix \ref{apthree},
if the strength of the
disorder is equal to 0, the bosonic and fermionic sectors decouple and then
the quantity $H$ will coincide with the purely bosonic correlator
\be
2 \VEV{\Theta^b(x) \Theta^b(0)} = {H(z \bar z) \over z^2 \bar z^2}
,\ee
where $\Theta^b$ refers to the  bosonic subsector of the theory. Since
the bosonic subsector is unitary, $H \ge 0$. 
As we increase the strength of the
disorder $H$ may change and in principle can change its sign. Apparently
it maintains its sign for small disorder, unless it is a discontinuous
function of disorder strength. We also note that $H$ may
change sign at the points along the renormalization group trajectory
where $\VEV{\Theta(x) \theta(0)} = 0$. 

At the critical point of the theory $H=0$ because the trace
of the energy momentum tensor $\Theta=0$. Then the parameter $b$ reaches
a fixed value, which we will also call $b$. 
What is the meaning of $b$ at the critical point? Obviously
it coincides with the correlator
\be
\label{bdef}
\VEV{T(z) t(0)} = {b \over z^4}
.\ee
By analogy with
appendix \ref{apthree}, consider a critical disordered
theory
with   the strength of the disorder equal to zero. For example, 
we can take the theory \rf{freeE} and set $E=0$ to make it critical. 
Just as in appendix \ref{apthree}, 
$T=T^b+T^f$ and $t=T^b-T^f$ where $T^b$ is the energy momentum tensor
of the bosonic sector and $T^f$ is that of the fermionic sector. 
Then it is obvious that
$b=c$
where $c$ is the central charge of the bosonic sector.

Thus $b$ coincides with the central charge of the bosonic theory
in the absence of disorder, and it becomes a genuine new parameter
as we turn on disorder. Figuratively speaking we can say that
while the central charge of the
supersymmetric theory counts the number of bosonic degrees of freedom
minus fermionic degrees of freedom, which is always 0, $b$ on the other hand
counts the number of bosonic plus the number of fermionic degrees of
freedom (which of course is just 
twice the number of bosonic degrees of freedom). 

What does Eq. \rf{bdef} imply for the conformal theory of the critical
point? Since $t_{\mu \nu}$ is not necessarily conserved tensor, $t$ does
not have to be holomorphic. However, assuming that it is, 
we are led to propose the existence of the
following algebra for the expansion modes of $T=\sum_n L_{n} z^{-n-2}$
and $t=\sum_n l_{n} z^{-n-2}$
\bea
\label{genV}
[L_{n}, L_{m} ] = (n-m) L_{n+m} \br
[L_{n}, l_{m} ] = (n-m) l_{n+m} + {b \over 6} (n^3-n) \delta_{n, -m} \br
[l_{n}, l_{m} ] = (n-m) L_{n+m}
\eea
which generalizes the Virasoro algebra for the critical points of
two dimensional disordered systems. The absence of the central extension
in the last commutator in \rf{genV} stems from the fact that $t$ is
defined up to addition of $T$. 

The algebra \rf{genV} is certainly present for the
Kac-Moody algebras \ct{KZ} 
based on the supergroups \ct{CMW} because $t$ is going to be
quadratic in currents
and therefore conserved. We compute the parameter $b$ for them in the
appendix \ref{apDirac}. 

Does the algebra \rf{genV} have to be  present for other more generic
disordered critical points? The answer to this question depends crucially
on whether $t$ is holomorphic at those points. While we lack a general
proof that $t$ has to be holomorphic, we can show that in order for the
primary operators of this theory to have nonzero correlation functions
the holomorphic operator
$t$ has to appear on the right hand side of various operator product
expansions. 

Indeed, consider a primary operator $V$. The general rules of conformal
field theory \ct{BPZ} dictate the operator product expansion of $V$
with itself,
\be
\label{curo}
V(z) V(0) = {C \over z^{2 h_V}} \left( {c \over h_V} + 2 z^2 T(0) + \dots
\right) + \dots
\ee
where $C$ is an overall normalization constant, $h_V$ is the dimension of
$V$ and $c$ is the central charge (see also \ct{Zamcur}). 
It follows from \rf{curo} that the
correlation function $\VEV{V(z) V(0)}$ is proportional to the central charge
of the theory,
\be
\VEV{V(z) V(0)} = C {c/h_V \over z^{2 h_V}}
.\ee
Normally the normalization constant $C$ is chosen to be inversely proportional
to the central charge to cancel that dependence. However,
for the theories considered in this paper $c=0$ and it looks like
all the correlation
function of these theories have to be zero! One
way
this paradox can be resolved is
by including a holomorphic field $t$ in the operator product expansion
of $V$ with itself,
\be
V(z) V(0)  = {C \over z^{2 h_V}} \left( {2 b \over h_V} + 2 z^2 t(0) + \dots
\right) + \dots
\ee
because while $L_{2} T=0$, $L_2 t = b$ as follows from \rf{genV}. This
is exactly what happens with the current algebra considered in the
appendix \ref{apDirac}. 

The situation may become more complicated if we are led to include
zero dimensional 
logarithmic operators \ct{ya} which have unusual expectation values. We
are not going to consider this possibility here. 

Now what is the physical meaning of $b$? According to \rf{genV}
$b$ is related to the transformation properties of $t$ under conformal
transformations. This will allow us to show that $b$ relates the physical
free
energy of the disordered system on a cylinder to its circumference $L$
as in the following,
\be
\label{freenergy}
F_{\rm phys} =-{\pi b \over 6 L}
\ee

Indeed, it is well known \ct{Affleck} that
the central charge $c$ of a conformal field theory relates the free energy
of the theory on the cylinder to the circumference of the cylinder $L$,
\be
\label{freenormal}
F=- {\pi c \over 6 L}
\ee
The central charge of our theory is $0$, which is only natural because the
free energy of this theory is also zero, being derived from the partition
function which is equal to $1$. Nevertheless, there is a well defined
physical free energy of the system, which can be computed as a disorder
averaged logarithm of the physical partition function defined without
using supersymmetry or replica tricks. It is this free energy
that enters \rf{freenergy}.

To show that \rf{freenergy} holds, 
we use that the expectation value of the tensor $t_{\mu \nu}$
of the supersymmetric theory is related to the expectation value of the
averaged over disorder physical energy momentum tensor,
\be
\label{physical}
\VEV{t_{\mu \nu}} = 2 \VEV {T^{\rm phys}_{\mu \nu}}
.\ee
This relation is discussed in appendix \ref{apfour} where  a careful
definition of the physical free energy is given. Note that a naive
generalization of this relation to higher order correlators 
would not be correct. 

On the other hand, the expectation value of $t$ computed on the cylinder 
with circumference $L$ is
given, by standard arguments, 
\be
\VEV{t_{\rm cyl}} = - {b \pi^2 \over 3 L^2}
\ee
because it follows from \rf{genV} that 
the transformation law for $t$ coincides with the transformation
law of the energy momentum tensor in the conformal field theory with
$c=2b$.

From now on, we can easily derive \rf{freenergy} by essentially
following Affleck's derivation of \rf{freenormal} in \ct{Affleck}
step by step and combining it with \rf{physical}. 

Therefore we see that the parameter $b$ is a way for the supersymetric
system to remember its nonsupersymmetric disordered `past'.

The author is grateful to A. Ludwig and C. Nayak for many stimulating
discussions and to C. Mudry for valuable comments on the manuscript. 
This work has been supported by the NSF grant PHY
94-07194.

\appendix
\section{}
\label{apone}
Consider the supersymmetric action $S$ describing the
disordered system before the average over disorder has been carried out. 
\be
\label{action}
S=\int \phi^* (E-H) \phi + \psi^* (E-H) \psi
\ee
where $H$ is a hamiltonian of a disordered system, including disorder,
$E$ is a parameter (energy), $\phi$ are commuting and $\psi$ are 
anticommuting variables. 
A basic property of this action is its invariance under the supergroup
U$(1|1)$. This group has four generators, $J$, $j$, $\eta$, and $\bar \eta$,
which act on the basic fields $\psi$ and $\phi$ in the following way
\ct{CMW}
\bea
\label{rules}
J: \, \delta \phi = \lambda \phi, \, \delta \phi^*= - \lambda \phi^*, \,
\delta \psi = \lambda \psi, \, \delta \psi^* = - \lambda \psi^*
\br
j: \, \delta \phi = \lambda \phi, \, \delta \phi^*= - \lambda \phi^*, \,
\delta \psi = - \lambda \psi, \, \delta \psi^* =  \lambda \psi^*
\br
\eta: \delta \phi= \epsilon \psi, \, \delta \psi^* = - \epsilon \phi^*
\br
\bar \eta: \delta \phi^* = \epsilon \psi^*, \, \delta \psi = \epsilon \phi
\eea
where $\lambda$ is a commuting and $\epsilon$ an anticommuting parameter. 

These generators obey the following commutation relations
\bea
\label{comm}
[j, \eta] = 2 \eta, \, [j, \bar \eta] = - 2 \bar \eta \br
\{ \eta , \bar \eta \} = J
\eea
with $[\,]$ being the commutator and $\{\,\}$ being the anticommutator. All
other commutators in this algebra are trivial.

A remarkable property of the action $S$ is that it can be expressed
as a variation of some other quantity under either $\eta$ or
$\bar \eta$,
\be
S= - \delta_{\eta} \left( \int \psi^* (E-H) \phi \right) = \delta_{\bar \eta}
\left( \int \phi^* (E-H) \psi \right)
\ee 
This property remains to be true even after the average with respect to
disorder is taken \ct{Bernard}. Moreover, the action $S$ is just a part
of a supermultiplet. The other members of the supermultiplet
are 
\bea
\zeta = \int \phi^* (E-H) \psi \br
\bar \zeta = - \int \psi^* (E-H) \phi \br
s = \int \phi^* (E-H) \phi - \psi^* (E-H) \psi
\eea
which transform under U$(1|1)$ as $\eta$, $\bar \eta$, and $j$.
The action itself transforms
as $J$, that is, it does not transform under U$(1|1)$, as in 
\rf{comm}. We also note that after averaging over disorder
$S$, $\zeta$, $\bar \zeta$, and $s$ become more complicated nonquadratic
functions of the basic fields $\phi$ and $\psi$. 
Nevertheless they preserve their transformation
properties. 

Analogously if $T_{\mu \nu}$ 
is an energy momentum tensor of the supersymmetric theory
of this kind, it is also part of the supermultiplet and transforms  as $J$.
The remaining tensors $\xi_{\mu \nu}$, $\bar \xi_{\mu \nu}$ and
$t_{\mu \nu}$ transform as $\eta$, $\bar \eta$ and $j$.

\section{}
\label{aptwo}

Let us prove that
\be
\label{impr}
\VEV{T_{\mu \nu}(x) t_{\mu' \nu'}(0)} = \VEV {t_{\mu \nu}(x) T_{\mu' \nu'}(0)}
\ee

Consider the following correlation function
\be
\VEV{t_{\mu \nu}(x) \xi_{\mu' \nu'}(0)}
\ee
By computing its variation with respect to $\bar \eta$ we
obtain
\be
\label{one}
- 2 \VEV{\bar \xi_{\mu \nu}(x) \xi_{\mu' \nu'}(0) } +
\VEV{t_{\mu \nu}(x) T_{\mu' \nu'}(0)} = 0
\ee
Analogously, by computing the variation of $\VEV{\bar \xi_{\mu \nu} (x)
t_{\mu' \nu'}(0)}$ under $\eta$ we get
\be
\label{two}
\VEV{ T_{\mu \nu}(x) t_{\mu' \nu'}(0)} - 2 \VEV{\bar \xi_{\mu \nu} (x)
\xi_{\mu' \nu'}(0)} =0 
\ee
Combining \rf{one} and \rf{two} we arrive at \rf{impr}.

\section{}
\label{apthree}

Consider a field theory corresponding to the disordered hamiltonian
\be
H=- \Delta + V(x)
\ee
where we set the disorder strength equal to zero. It is given by
\be
\label{freeE}
S=\int \phi^* (\Delta + E) \phi + \psi^* (\Delta+E) \psi
\ee
This field theory is free and easily solvable. 
The total energy momentum tensor is a sum of the energy momentum
tensors of the bosonic and fermionic sectors of the theory,
\be
T_{\mu \nu} = T^b_{\mu \nu} + T^f_{\mu \nu}
\ee
As we discussed before, all the correlations of the energy
momentum tensor with itself will be equal to zero because of the cancellation
between fermionic and bosonic correlators. For example,
\be
\VEV{T_{\mu \nu}(x) T_{\mu' \nu'}(0)} = \VEV{T^b_{\mu \nu}(x)
T^b_{\mu' \nu'} (0)}
+ \VEV{T^f_{\mu \nu}(x) T^f_{\mu' \nu'}(0)} = 0
\ee
On the other hand, let us define the tensor $t_{\mu \nu}$ to be
\be
t_{\mu \nu}= T^b_{\mu \nu} - T^f_{\mu \nu}
\ee
It is not difficult to check that it coincides with the tensor
$t$ as defined in appendix 1 by computing its transformation
properties under U$(1|1)$. 
Its correlator with $T$ is equal, on the other hand,  to
\be
\VEV{T_{\mu \nu}(x) t_{\mu' \nu'}(0)} = \VEV{T^b_{\mu \nu}(x)
T^b_{\mu' \nu'} (0)}
- \VEV{T^f_{\mu \nu}(x) T^f_{\mu' \nu'}(0)} = 2  \VEV{T^b_{\mu \nu}(x)
T^b_{\mu' \nu'} (0)}
\ee
Thus it reproduces the correlator of the purely bosonic (unitary)
theory!

\section{}
\label{apDirac}

Let us determine $b$ for U$(1|1)$ current algebra.
It was studied in \ct{CMW} where it was shown that
it describes the two dimensional
Dirac
fermions in the presence of random gauge potential. 
This algebra depends on two parameters, central extensions
$k$ and $k_j$. For random Dirac
fermions $k=1$ while $k_j$ could be arbitrary positive real number
related to disorder
strength. More generally, $k$ is allowed to be $1$ over 
an arbitrary integer number.
However, the physical meaning of that algebra with $k<1$ is
not completely understood in the literature. 

The energy momentum tensor of such a theory is
quadratic in the U$(1|1)$ currents $J$, $j$, $\eta$, and
$\bar \eta$, and is given by
\be
T={1 \over 2 k} \left( J j + \eta \bar \eta - \bar \eta \eta \right)
+{ 4 - k_j \over 8 k^2} JJ
\ee
Of course, the central charge of this conformal field theory is equal to
0, as could be checked by computing $\VEV{T(z) T(0)}$ \ct{CMW}. The 
parameter $b$ could, on the other hand, be computed in the following way.

First we find the other components of the supermultiplet $T$, $t$,
$\xi$, and $\bar \xi$, by constructing expressions quadratic in currents
which obey the right commutation relations, 
\bea
\xi = {1 \over 4 k} \left( \eta j + j \eta \right) + { 4 - k_j \over 8 k^2}
\eta J \br
T= \{ \bar \eta \, , \xi \}
\eea
and
\bea
t={1 \over 4 k} j j + { 4 - k_j \over 16 k^2} \left( Jj + \bar \eta
\eta - \eta \bar \eta \right)
\br
2 \xi = [ \eta \, , t]
\eea
Note that $t$ could only be found up to addition of other primary
scalar fields of dimension 2, such as $T$ or $JJ$. This ambiguity does
not affect $b$, however.

By computing the correlation function $\VEV{T(z) t(0)}$ with the help
of the operator product expansion of the U$(1|1)$ currents taken 
from \ct{CMW}
we
can determine $b$ which turns out to be 
\be
b = {1 \over k}
\ee

So indeed $b$ has all the properties we might have intuitively
expected.  
When the disorder strength $k_j=0$, $b$ coincides with the central
charge of nonrandom Dirac fermions, $c=1$. As the disorder strength increases
$b$ does not change. This is in agreement with the theorem proved in this
paper, because the theory is critical for all values of $k_j$ and
$H$ has to vanish everywhere. Thus $b$ cannot change according to 
\rf{ctheorem}. 

For other critical points $b=m$ where $k=1/m$, $m$ being an arbitrary 
integer.

We note that $t$ is automatically holomorphic in this construction.
Therefore, $l_n$ and $L_n$ form the algebra \rf{genV}. 

\section{}
\label{apfour}

We want to show that 
\be
\label{physicl}
\VEV{t_{\mu \nu}} = 2 \VEV {T^{\rm phys}_{\mu \nu}}
\ee

First of all, let us define what we mean by the physical free energy. 
The physical partition function for the disordered system with random
Hamiltonian $H$ is given by
\be
Z=\int {\cal D} \phi^* {\cal D} \phi  \, e^{- S^b}
\ee
where $S^b$ is a bosonic part of \rf{action}, $S^b=\int \phi^* (H-E) \phi$
while the free energy is given by the disorder averaged logarithm of $Z$,
\be
F_{\rm phys} = \VEV{ \log Z}_{\rm disorder}
\ee
Suppose now that the theory is defined on curved manifold with a metric
$g_{\mu \nu}$. Then the expectation value of the physical energy momentum
tensor can be found by differentiating $F_{\rm phys}$ with respect to 
$g_{\mu \nu}$,
\be
- { \delta F_{\rm phys} \over \delta g_{\mu \nu}} = 
\VEV{ {\delta S \over \delta g_{\mu \nu}}
\over Z}_{\rm disorder} \equiv \VEV{T^{\rm phys}_{\mu \nu}}_{\rm disorder}
\ee
Now the $1/Z$ term can be reexpressed as a fermionic path integral,
\be
{1 \over Z} = \int {\cal D} \psi^* {\cal D} \psi \,  e^{- S^f}
\ee
with $S^f$ being the fermionic part of the action \rf{action}. 

Therefore, we arrive at the following formula, 
\be
- { \delta F_{\rm phys} \over \delta g_{\mu \nu}} = \VEV{ 
\int {\cal D} [\psi, \psi^*, \phi, \phi^*] \, {\delta S^b \over 
\delta g_{\mu \nu}} \, e^{-S}}_{\rm disorder}
\ee
with $S=S^b+S^f$. 

Now it is not hard to see that 
\be 
\label{impro}
\VEV{t_{\mu \nu}} = \VEV{t_{\mu \nu}+ T_{\mu \nu}} =
2 \VEV{ 
\int {\cal D} [\psi, \psi^*, \phi, \phi^*] \, {\delta S^b \over 
\delta g_{\mu \nu}} \, e^{-S}}_{\rm disorder} 
\ee
We have used that $\VEV{T_{\mu \nu}} = 0$ and the explicit construction
of appendix \ref{apone}. \rf{physicl} obviously follows.

\end{document}